\documentclass[twocolumn]{aastex631}

\usepackage{xspace}

\newcommand{\tardishe}{\textsc{tardis-he}\xspace}
\usepackage{hyperref}
\usepackage{listings}
\usepackage{amsmath}
\hypersetup{colorlinks,linkcolor={magenta},citecolor={cyan},urlcolor={blue}}  

\graphicspath{{./}{figures/}}

\begin{document}

\title{Effect of positronium on the $\gamma$-ray spectra and energy deposition in Type Ia supernovae}

\author[0000-0002-7708-3831]{Anirban Dutta}
\affiliation{Department of Physics and Astronomy, Michigan State University, East Lansing, MI 48824, USA}

\author[0000-0001-7343-1678]{Andrew Fullard}
\affiliation{Department of Physics and Astronomy, Michigan State University, East Lansing, MI 48824, USA}

\author[0000-0002-0479-7235]{Wolfgang Kerzendorf}
\affiliation{Department of Physics and Astronomy, Michigan State University, East Lansing, MI 48824, USA}

\author[0000-0003-3615-9593]{J. T. O'Brien}
\affiliation{Department of Astronomy University of Illinois Urbana-Champaign, Champaign, IL 61801-3633}

\author[0009-0007-6332-2188]{Cecelia Powers}
\affiliation{Department of Physics and Astronomy, Michigan State University, East Lansing, MI 48824, USA}

\author[0000-0002-9774-1192]{Stuart A Sim}
\affiliation{Astrophysics Research Centre, School of Mathematics and Physics, Queen's University Belfast, Belfast BT7 1NN, Northern Ireland, UK}

\author[0000-0003-2024-2819]{Andreas Flörs}
\affiliation{GSI Helmholtzzentrum für Schwerionenforschung, Planckstraße 1, D-64291 Darmstadt, Germany}

\author[0000-0002-4391-6137]{Or Graur}
\affiliation{Institute of Cosmology and Gravitation, University of Portsmouth, Portsmouth PO1 3FX, UK}
\affiliation{Department of Astrophysics, American Museum of Natural History, Central Park West and 79th Street, New York, NY 10024-5192, USA}

\begin{abstract}

Type Ia supernovae (SNe Ia) are powered by the radioactive decay of isotopes such as $^{56}$Ni and $^{56}$Co, making their $\gamma$-ray spectra useful probes of the explosion mechanism and ejecta structure. Accurate interpretation of $\gamma$-ray observables, including line ratios and continuum fluxes, requires a detailed understanding of the microphysical processes that shape the spectra. One such process is positronium formation during electron-positron annihilation, which can redistribute flux from the 511 keV line into the surrounding continuum. To assess the impact of positronium on the emergent spectra, we developed a new open-source module, \tardishe, for time-dependent three-dimensional $\gamma$-ray transport, integrated into the radiative transfer code \textsc{tardis}. The code simulates $\gamma$-ray spectra and light curves from one-dimensional supernova ejecta models and allows for flexible incorporation of decay chains and opacity treatments.
Using \tardishe, we explore the effect of positronium formation by varying the positronium fraction from 0\% to 100\%, and assuming an extreme case where 75\% of positronium decays result in three-photon emission. We find that full positronium formation can reduce the 511 keV line flux by $\approx$ 70\% and modestly enhance energy deposition by up to 2\% at around 100 days post-explosion, compared to models without positronium. These results demonstrate that while the effect is not dominant, positronium formation introduces measurable changes to $\gamma$-ray observables. Future observations with missions such as the Compton Spectrometer and Imager (COSI) may offer constraints on positronium formation in SNe Ia and help refine models of their radioactive energy transport.

\end{abstract}

\section{Introduction}

Type Ia supernovae (SNe Ia) are thermonuclear explosions of white dwarfs (WDs). These events play a crucial role in galactic chemical evolution \citep[see Figure~39 by][]{2020ApJ...900..179K}. The empirical relation between the decline rate of the light curve and the peak magnitude make SNe Ia one of the most precise distance indicators for cosmology \citep{1977SvA....21..675P, 1993ApJ...413L.105P}. 

Despite their central role in modern astrophysics, the progenitor, ignition, and explosion conditions remain a mystery and highly debated \citep[see e.g.][for recent reviews]{2025A&ARv..33....1R, 2019NatAs...3..706J, 2018PhR...736....1L} that result in significant uncertainties in the field that they are applied to. Almost all scenarios that are consistent with observations require the exploding WD to be in a binary with a variation in the mass of the exploding white dwarf, ignition mechanism, and flame propagation during the explosion. The models that have been successful in explaining some of the observed properties of SNe Ia are deflagration to detonation in a near $M_{ch}$ WD \citep[see for e.g.][]{1991A&A...245..114K, 1995ApJ...444..831H}, double detonation in a sub-$M_{ch}$ WD \citep[see][]{ 1994ApJ...423..371W, 1995ApJ...452...62L}, and merger of two WDs \citep{1984ApJ...277..355W, 2010Natur.463...61P}. There also exists peculiar class of WD explosions such as 2002cx-like \citep{2003PASP..115..453L, 2013ApJ...767...57F, 2022ApJ...925..217D} for which pure deflagration models have been proposed \citep{2006AJ....132..189J}.  Finally, \cite{2021PhRvL.126m1101H} studied isolated white dwarfs ignited by fission chain reactions that might also explain the phenomenon. 

All current explosion models for SNe Ia broadly reproduce the spectra and light curves in the UVOIR regime. \cite{2012ApJ...750L..19R} compared the delayed detonation \citep{2013MNRAS.429.1156S} and violent merger \citep{2012ApJ...747L..10P} models of SNe Ia to find that early and near-maximum spectral features do not offer clear distinction of one over another. 
While, spectra have the potential to serve as discriminants, their strong sensitivity to detailed chemical composition, density profiles, and ejecta temperature makes them unreliable for distinguishing between progenitor scenarios specfically near maximum. Even slight variations in the progenitor structure can lead to significantly different spectral features \citep{obrien_1991t-like_2024}.

SNe Ia are primarily powered by the radioactive decay of $^{56}$Ni and $^{56}$Co \citep{1962PhDT........25P, 1969ApJ...157..623C}. The smaller size of the progenitor and the production of large amounts of radioactive materials means more $\gamma$-rays can escape to be observed. The production of $\gamma$-rays and their subsequent transport are governed by well-known decay chains, branching ratios, and a few key interaction processes, including pair production, Compton scattering, and photoelectric absorption \citep{2004ApJ...613.1101M}. 
As a result, $\gamma$-ray observations provide a straightforward but powerful means of investigating the mass-velocity distribution of the ejecta and potentially serve as valuable diagnostics for distinguishing between different progenitor scenarios \citep{2013A&A...554A..67S}.

Several studies have shown that even in the early phase, where most $\gamma$-rays are trapped, the $\gamma$-ray line ratios relative to the continuum can be used to infer the composition and density-velocity distributions \citep{1990ApJ...360..626B, 1998ApJ...492..228H, 1998MNRAS.295....1G, 2008MNRAS.385.1681S}. Direct $\gamma$-ray observables are challenging to detect due to the limited number of nearby events, but SN~2014J in M82 provided a unique opportunity. The detection of $^{56}$Co lines at 847 keV
and 1238 keV energies supported the explosion of a white dwarf \citep{2014Natur.512..406C, 2015ApJ...812...62C}. The early observations of 158 keV and 812 keV $\gamma$-ray lines from $^{56}$Ni decay supported an explosion mechanism in which $^{56}$Ni is present in the outer layers of the SN ejecta to some extent \citep{2014Sci...345.1162D}. Indirect observations like the $\gamma$-ray escape timescales inferred from observed bolometric light curves and the time evolution of energy deposition provide alternative approach to probing explosion physics \citep{2019MNRAS.484.3941W, 2020MNRAS.496.4517S, 2023MNRAS.522.6264S, 2024MNRAS.535..924S}. As the ejecta become optically thin to the $\gamma$-rays, the measurement of line fluxes gives an estimate of the mass of the radioactive isotope produced in the explosion. \citet{2008MNRAS.385.1681S, 2013A&A...554A..67S} have investigated different line ratios and line-to-continuum ratio as diagnostics for the explosion models.

The 511 keV $\gamma$-ray line, arising from electron-positron annihilation, is often excluded from supernova $\gamma$-ray studies as it does not directly probe specific isotopes and decay channels. However, electron-positron pair annihilation can significantly affect $\gamma$-ray emission and associated observables. In particular, annihilation does not always proceed directly to produce two 511 keV photons. Instead, a positron can briefly form a bound state with an electron -- known as positronium -- before annihilating. This intermediate state alters the resulting spectrum: while direct annihilation contributes to the 511 keV line, formation of positronium and its decay produce a continuum below 511 keV. This positronium formation redistributes some of the expected line energy into the continuum, which can change the interpretation of observed line fluxes measured against the continuum. However, most studies do not include the positronium pathway for annihilation.

In this work, we introduce a new code \tardishe  to study the propagation of high-energy photons ($\gamma$-rays and $X$-rays) and include the positronium effect. We specifically focus on what effect the positronium channel has on spectra and deposition curves. 

The paper is organized as follows. In Section~\ref{methods}, we discuss the methodology and the implementation of the high energy photon transfer within the SN ejecta. Then we compare the $\gamma$-ray spectra and the energy deposited by the $\gamma$-rays and positrons from \tardishe to results from the literature and other codes in Section~\ref{code_compare}. In Section \ref{results}, we discuss the effect of positronium on the $\gamma$-ray observables and the deposition curve. We also apply the code to different SNe Ia models to find features that can discriminate between explosion scenarios. We conclude with discussion on positronium formation.

\section{Method} \label{methods}
\tardishe is based on the Monte Carlo radiative transfer methods outlined in \citet{1983ASPRv...2..189P, 1988ApJ...325..820A} and \citet{2005A&A...429...19L}.  The code uses indivisible energy packets  to represent the radiation field. Each energy packet is made up of photons with the same frequency. In the indivisible packet scheme, the frequency of the photons in the packet and the location/direction of the packet \citep{2005A&A...429...19L} can change, but the packets are never split - that is, the energy of the packet remains the same in the comoving frame. Each packet is initialized and its propagation is followed as it performs different interactions within each simulation cell at each time step. Finally, the time-dependent $\gamma$-ray spectra and energy depositions are calculated. 

The simulation volume is configured to be multiple spherical shells characterized by density, velocity, and mass fractions of the different elements and isotopes \citep[see][]{2014MNRAS.440..387K}. The energy packets are initialized based on the radioactive decay energy and the simulation tracks their time-dependent transport through the ejecta.

\subsection{Radioactive decay and radiation}

The radioactive decay of the ejecta impacts the composition as well as the energy generation in the ejecta. We first calculate the masses of the isotopes in each shell. The \textsc{radioactivedecay} package \citep{Malins2022} uses the masses to calculate the total number of decays. Changes in compositions for each time step as the isotopes decay are also taken into account.

We use the decay radiation and transition probabilities from the National Nuclear Data Center \footnote{\url{https://www.nndc.bnl.gov}} in the form of an Evaluated Nuclear Structure Data File (\texttt{ENSDF})\footnote{\url{https://www.nndc.bnl.gov/ensdfarchivals/}} \citep{2011NDS...112.1513J}. Specifically, we use the \textsc{ensdf$\_$230601} dataset \footnote{\url{10.18139/nndc.ensdf/1845010}} prepared with the \textsc{carsus} \footnote{\url{https://tardis-sn.github.io/carsus/installation.html}} package which manages atomic and nuclear datasets. 

The decay from one isotope to another is divided into various transition channels \citep[simplified decay schemes for $^{56}$Ni and $^{56}$Co are given by][]{1994ApJS...92..527N}. The ENSDF files provide an energy ($E_{l}$) and an associated probability ($f_{l}$) per 100 decays from an energy level \citep[using the notation from][]{2005A&A...429...19L} for each transition. Thus, the energy per decay of that particular line is $E_{l} f_{l}$, and the total energy per decay of the isotope from all the transitions is $\sum E_{l}f_{l}$. We compile a table that contains the decay-radiation types and energies for each channel, for each time step ($\Delta t_i$), and each shell $k$. Table~\ref{tab:radiation_energy} shows an excerpt of the full table. 

We calculate the energy from $N_\textrm{decay, k}(\Delta t_i)$ for each channel and each time step $\Delta t_i$ by multiplying the total number of decays ($N_\textrm{decay, k}$) with the energy of each channel ($E_{l}f_{l}$). 
\begin{equation}
    E_{\textrm{decay,\ k}}(\Delta t_i) = N_{\rm decay,\ k}(\Delta t_i) E_{l} f_{l}
\end{equation}

We calculate the total energy of the decay radiation for channels that decay via electromagnetic radiation and for $\beta^{+}$-decay channels $(E_{\textrm{total, }\beta^{+}})$ across the entire simulation (for all time-steps):
\begin{align}
E_\textrm{total, EM} &= \sum_{i, k, \textrm{EM}} E_{\textrm{decay, EM},\ k, \ i}\\
E_{\textrm{total},\ \beta^{+}} &= \sum_{i,k, \beta^{+}} E_{\textrm{decay},\ \beta^{+}, \ k,\ i}
\end{align}

The kinetic energy of the $\beta^{+}$ -- decay channels is assumed to be immediately deposited in the ejecta as thermal energy. Energy from electromagnetic radiation is split up into packets used in the Monte Carlo transport. 

\subsection{High Energy Monte Carlo packet initialization}
The high energy Monte Carlo packets have the following properties - rest-frame frequency ($\nu_\textrm{rf}$), rest-frame energy ($E_\textrm{rf}$), time of propagation ($t_\textrm{propagation}$), direction ($\theta, \phi $), and location ($r$). 

For $N_\textrm{packets}$, we sample from the entire electro-magnetic decay radiation table (excerpt given in Table~\ref{tab:decay radiation}) weighted by $E_{\textrm{decay}, k}(\Delta t_i)$. A sampled packet has a specific decay radiation channel, a specific time step $\Delta t_i$, and a specific shell $k$. We chose to set $t_\textrm{propagation}$ to the start of the time step that was sampled.

We calculate the initial radial location for the packet by first sampling the velocity location in the ejecta 
\begin{equation}
v = [zv_{k, \textrm{inner}}^{3} + (1 - z)v_{k,\textrm{ outer}}^{3}]^{\frac{1}{3}}
\end{equation}
for the $k$-th shell, where $v_{k, \textrm{inner}}/v_{k, \textrm{outer}}$ are the velocities of the inner and outer boundaries of a shell and $z$ is a random number between $[0, 1)$. \tardishe assumes homology and thus we calculate the radial location of the packets $r=v t_\textrm{propagation}$. 

The directional properties of the packet are described by two angles. The polar angle (\( \theta \)) is sampled between \([0, \pi)\), while the azimuthal angle (\( \phi \)) is sampled between \([0, 2\pi)\) using random numbers as given by \citep[see][]{osti_4167844}:
\begin{equation}
\begin{split}
    \cos\theta = 1 - 2z_{1} \\
    \phi = 2 \pi z_{2}
\end{split}
\end{equation}

Each channel has a photon frequency $\nu_\textrm{channel}$. We apply a non-relativistic frame transformation to convert to the rest-frame frequency

\begin{equation}
    \nu_\textrm{rf} = \nu_\textrm{channel} / (1 - \frac{\vec{v} \cdot \hat{n}}{c}),
\end{equation}

Finally, we set the rest frame energy of the packet to
\begin{equation}
    E_\textrm{rf}=\frac{E_\textrm{total, EM}}{N_\textrm{packet}} / ( 1 - \frac{\vec{v} \cdot \hat{n}}{c})
\end{equation}
where \( \hat{n} \) is the direction vector of the packet and \( \vec{v} \) is the velocity of the ejecta at the position of the packet.

An exception is made if the sampled channel is the electron/positron annihilation line where positronium formation is treated.

\subsection{Initializing packets from positron annihilation}

The positrons generated from a decay have a chance to annihilate with electrons in the ejecta. Before annihilation, the positrons can form a bound state with an electron known as positronium \citep[Ps][]{https://doi.org/10.1002/asna.19342530402, doi:10.1142/S0217751X04020142}. Positronium can be formed in singlet (para-Ps) or triplet (ortho-Ps) spin states in the statistical ratio of 1\ $:$\ 3 \citep{1949PhRv...75.1696O}. Positronium decays by the emission of two $\gamma$-rays (2$\gamma$, para-Ps) or three $\gamma$-ray photons (3$\gamma$, ortho-Ps) in the ratio of 1\ $:$ \ 4.5 \citep{1976ApJ...210..582C} with different lifetimes. The 3$\gamma$-decay contributes to the continuum of the spectrum as the three $\gamma$-ray photons share the 1022 keV annihilation energy with a maximum energy of 511 keV. The 2$\gamma$-decay contributes to the 511 keV line. In this work, we only consider positronium formation due to positrons from radioactive decay and neglect the effect due to pair creation.

We simulate the contribution of positronium in our code by performing a separate treatment on packets that are generated in the annihilation channels. We choose a positronium fraction  ($f_{p}$) that is the fraction of annihilation packets (with a packet frequency of 511 keV/$h$) that undergo positronium formation \citep[see][for alternate definitions]{1987ApJ...319..637B}  before annihilation. For a non-zero positronium fraction we first randomly select annihilation packets until the given fraction is reached. For those that form positronium, 75 $\%$ decays to 3$\gamma$ to form a continuum. The 3$\gamma$ decay probability of 75 $\%$ is as given for laboratory measurements \citep{1987ApJ...323..159L, 2004ApJ...613.1101M}.
For the two-photon decay, the packets have a frequency of 511 keV/$h$. For the three-photon decay, we sample from the three-photon continuum using the method described in Appendix~\ref{three photon decay}.

\begin{figure*}
\centering
\includegraphics[width=\linewidth]{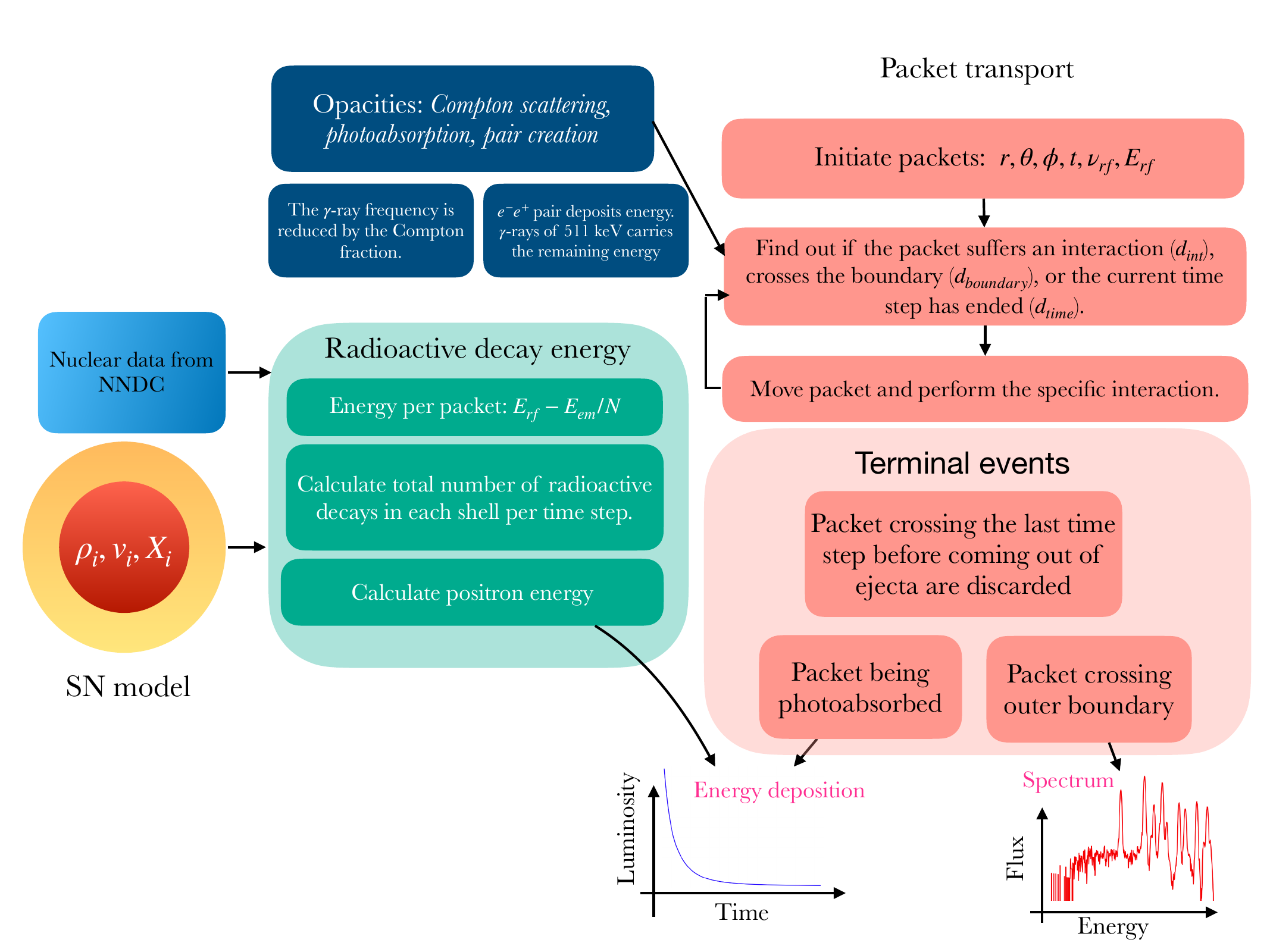}
\caption{Schematic diagram of \tardishe}
\label{fig:figure1}
\end{figure*}

\begin{table*}
\centering
\setlength{\tabcolsep}{3pt}
\caption{Decay energy from $^{56}$Ni}
\label{tab:decay radiation}
\begin{tabular}{c c c c c l c}
\hline \hline
  $t_{\rm start}$ & shell & number of decays ($^{*}$) & energy  & decay intensity & energy per decay  & decay energy  \\
  (days)  & ($k$)   &   & ($E_{l}$, keV)  &  ($f_{l}$) &   ($E_{l}f_{l}$ / 100, keV)   &  (keV) ($^{**}$)\\

\noalign{\smallskip} \hline \hline
2.0   &  0   &  1.24 $\times$ 10$^{50}$ & 158.38 & 98.8  & 156.48 & 1.93 $\times$ 10$^{52}$ \\
      &      &  1.24 $\times$ 10$^{50}$ & 811.85 & 86.0  & 698.19 & 8.62 $\times$ 10$^{52}$ \\ 
      &      & 1.24 $\times$ 10$^{50}$ & 749.95 & 49.5 & 371.23 & 4.59 $\times$ 10$^{52}$ \\ 
 \ldots &  & &  & \ldots & &  \ldots \\ 
      &  1   & 2.22 $\times$ 10$^{50}$ & 158.38 & 98.8 & 156.48 & 3.47 $\times$ 10$^{52}$ \\
      &      & 2.22 $\times$ 10$^{50}$ & 811.85 & 86.0 & 698.19 & 1.55 $\times$ 10$^{53}$ \\
      &      & 2.22 $\times$ 10$^{50}$ & 749.95 & 49.5 & 371.23 & 8.24 $\times$ 10$^{52}$ \\
 \ldots &  & &  & \ldots & &  \ldots \\
 \ldots &  & &  & \ldots & &  \ldots \\
2.09  &  0  & 1.28 $\times$ 10$^{50}$ & 158.38 & 98.8  & 156.48 & 2.00 $\times$ 10$^{52}$ \\
      &    &  1.28 $\times$ 10$^{50}$ & 811.85 & 86.0  & 698.19 & 8.93 $\times$ 10$^{52}$ \\
      &    & 1.28 $\times$ 10$^{50}$ & 749.95 & 49.5 & 371.23 & 8.23 $\times$ 10$^{52}$ \\
 \ldots &  & &  & \ldots & &  \ldots \\
      &  1  & 2.30 $\times$ 10$^{50}$ & 158.38 & 98.8 & 156.48 & 3.60 $\times$ 10$^{52}$ \\
      &     & 2.30 $\times$ 10$^{50}$ & 811.85 & 86.0 & 698.19 &  1.60 $\times$ 10$^{53}$ \\
      &     & 2.30 $\times$ 10$^{50}$ & 749.95 & 49.5 & 371.23 & 8.53 $\times$ 10$^{52}$ \\
 \ldots &  & &  & \ldots & &  \ldots \\
 \ldots &  & &  & \ldots & &  \ldots \\
191.0 &  0  & 3.20 $\times$ 10$^{42}$ & 158.38 & 98.8 & 156.48 & 5.00 $\times$ 10$^{44}$ \\
        &    & 3.20 $\times$ 10$^{42}$ & 811.85 & 86.0 & 698.19 & 2.20 $\times$ 10$^{45}$ \\
        &    & 3.20 $\times$ 10$^{42}$  & 749.95 & 49.5 & 371.23 & 1.19 $\times$ 10$^{45}$ \\
     \ldots &  & &  & \ldots & &  \ldots \\
        &  1  & 5.74 $\times$ 10$^{42}$ & 158.38 & 98.8 & 156.48 & 8.98 $\times$ 10$^{44}$ \\
        &      & 5.74 $\times$ 10$^{42}$ & 811.85 & 86.0 & 698.19 & 4.00 $\times$ 10$^{45}$ \\
        &      & 5.74 $\times$ 10$^{42}$ & 749.95 & 49.5 & 371.23 & 2.13$\times$ 10$^{45}$ \\
 \ldots &  & &  & \ldots & &  \ldots \\
\noalign{\smallskip} \hline \hline
\end{tabular}
    \vspace{2mm} 
 \begin{minipage}{0.8\textwidth}
        {\small
        \begin{itemize}
            \item[$^{*}$] The number of decays from a total of 0.67 $M_{\odot}$ of $^{56}$Ni in the ejecta in different shells and at different time steps. We run the angle averaged \textit{ddt N100} model \citep{2013MNRAS.429.1156S} with 92 shells from 2 to 200 days and 500 time steps. We consider all the decay channels for each isotope but show only three $\gamma$-ray lines of $^{56}$Ni with high $f_{l}$ for representation. 
            \item[$^{**}$] The decay energy is calculated by multiplying the number of decays times the energy of each channel.
        \end{itemize}
        }
    \label{tab:radiation_energy}
    \end{minipage}

\end{table*}

\subsection{High energy photon opacities}

The current version of \tardishe treats three interactions of the packets with the ejecta materials: i) Compton scattering, ii) pair production, and iii) photoabsorption. We will describe the calculations of opacities for each one of them in the following section.

\paragraph{Compton scattering}
After interacting with a free or bound electron, the $\gamma$-ray photon loses its energy and emits in a new direction. For photons with no initial polarization, the cross-section is independent of the azimuthal angle and anisotropic only in the polar direction (see more in Appendix~\ref{Klein-Nishina Equation}). The cross-section is independent of temperature and ionization state of the ejecta. We consider, all the electrons (bound and free) contribute to the Compton scattering. This is valid since the $\gamma$-ray energies are much higher compared to the binding energies of the electron. For example, the electron binding energy for the K-shell electron of Fe is $\sim$ 7 keV \footnote{\url{https://xdb.lbl.gov/Section1/Table_1-1.pdf}}.  We used the following integrated attenuation coefficient for a photon that undergoes Compton scattering \citep{1995qtf..book.....W}.
\begin{equation}
\begin{split}
     \alpha_\textrm{Compton} &= n_e \frac{3}{4} \sigma_T \\
     &[\frac{1 + x} {x^3}[
    \frac{2x(1 + x)}{1 + 2x} - \ln(1 + 2x)]\\
    & + \frac{1}{2x} \ln(1 + 2x)
    - \frac{1 + 3x}{(1 + 2x)^2}],
\end{split}
\end{equation}
where $x = \frac{h\nu}{m_{e}c^{2}}$, and $n_{e}$ is the electron number density.

\begin{figure*}[htbp]
    \centering
    \begin{minipage}[t]{0.45\textwidth}
        \centering
        \includegraphics[width=\textwidth]{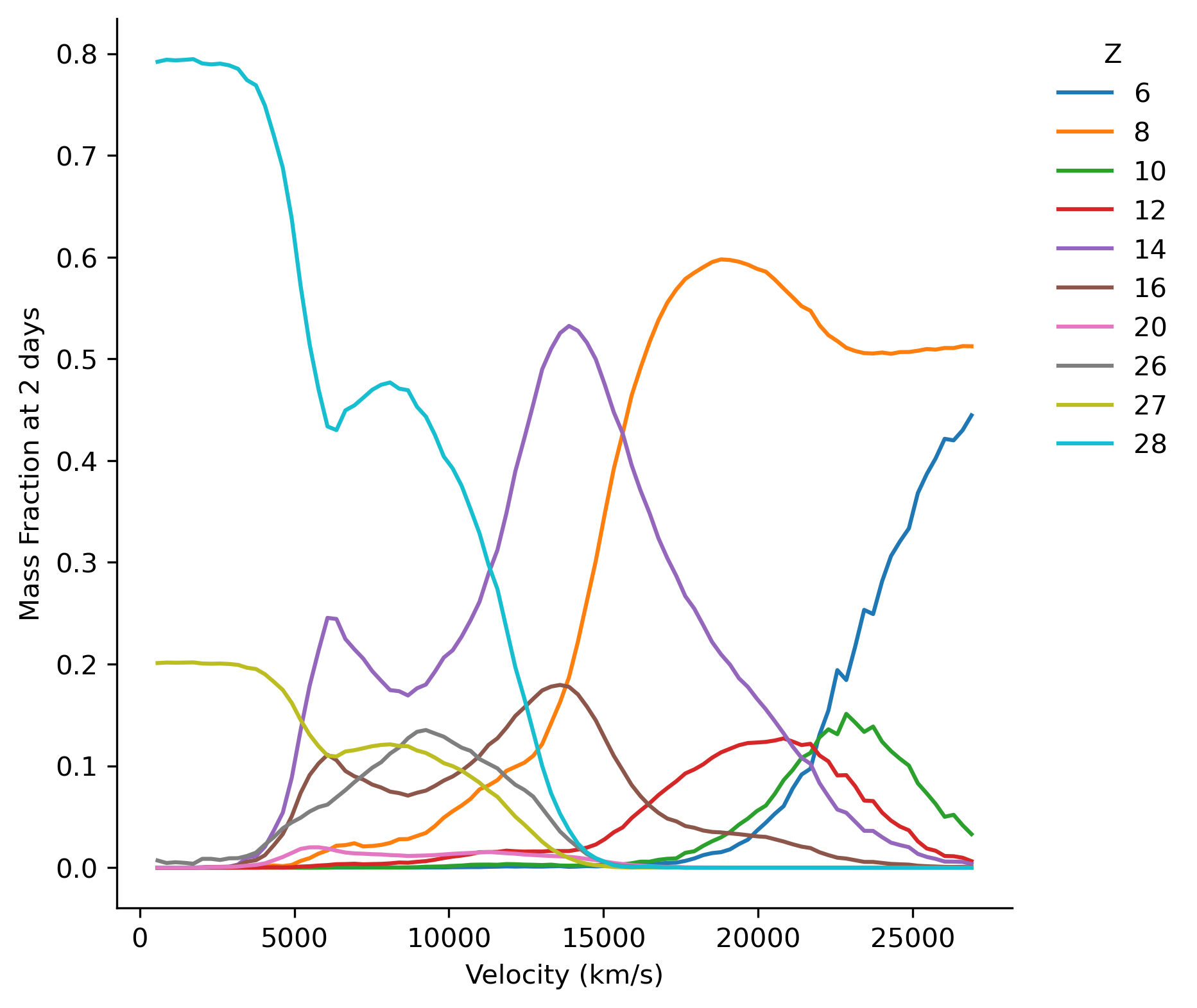}
        \caption{Mass fractions of the elements at 2 days for the \textit{ddt N100} model \citep{2013MNRAS.429.1156S}. Since the ejecta model is defined at 100 secs, the composition is decayed to the starting time of the simulation.}
        \label{fig:figure2}
    \end{minipage}
    \hfill 
    \begin{minipage}[t]{0.45\textwidth}
        \centering
        \includegraphics[width=\textwidth]{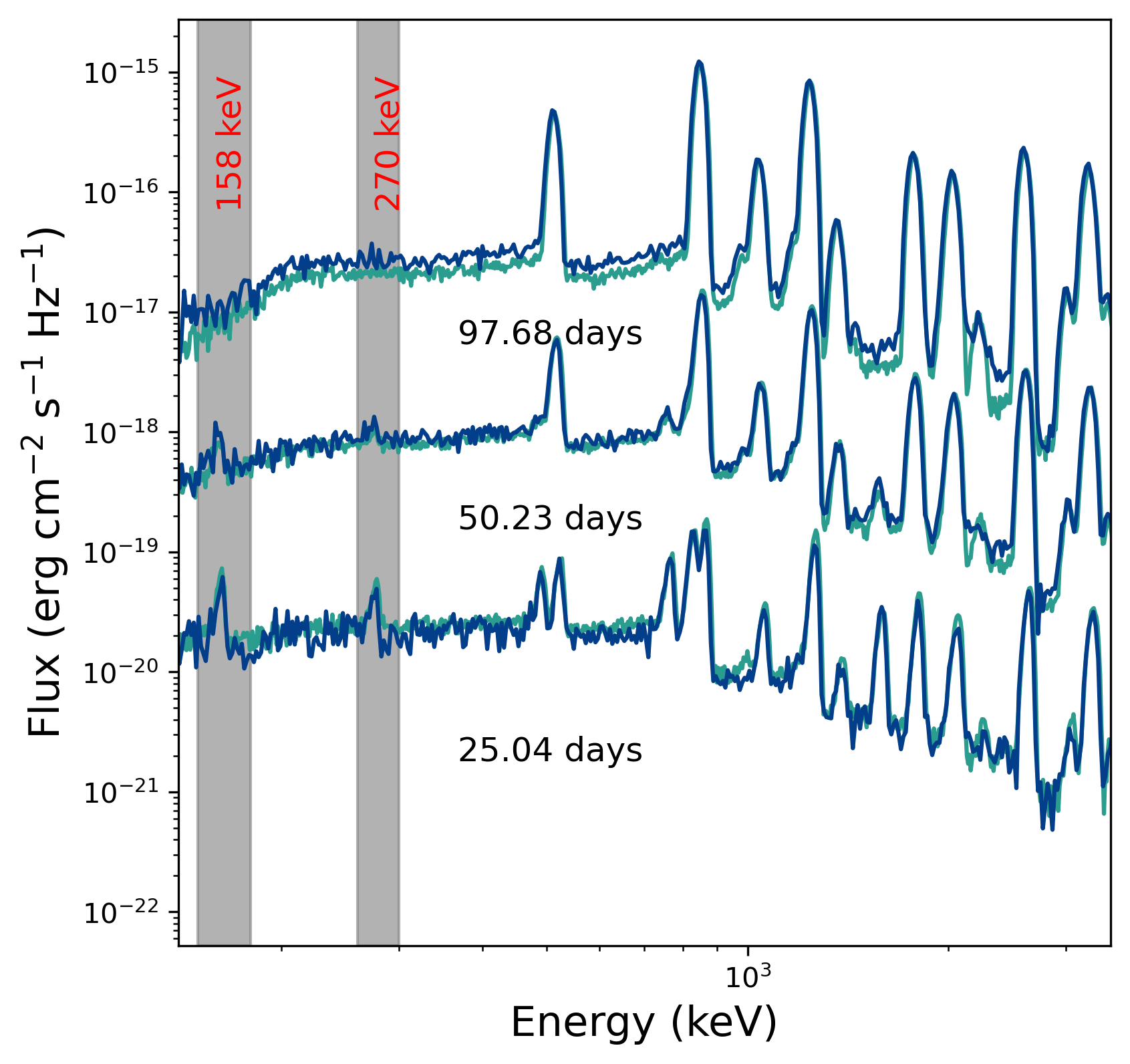}
        \caption{Comparison of \tardishe spectra (blue) at epochs indicated in the figure with ddt model from \cite{2013A&A...554A..67S} at similar epochs (green). We use 7 $\times$ 10$^{7}$ packets for our simulations. The grey regions  are the lines due to $^{56}$Ni at 158 keV and 270 keV which appears in the early phase and dissolves in the continuum as the SN evolves with time. The flux is calculated assuming a distance of 10 pc.}
        \label{fig:figure3}
    \end{minipage}
\end{figure*}

\paragraph{Pair creation}
While passing an atomic nucleus a $\gamma$-ray produces an $e^{-}$~$e^{+}$ pair. The electron will deposit its kinetic energy of $E_{\gamma}$/2 - $m_{e}c^{2}$, while a positron will release its kinetic energy and the annihilation energy of $E_{\gamma}$/2 - $m_{e}c^{2}$ + 2 $m_{e}c^{2}$.
We use the pair production attenuation coefficient ($\rm cm^{-1}$) given in \citet[][see Equation 2]{hubbell1969photon, 1988ApJ...325..820A}:
\begin{equation}
\begin{split}
\label{eqn:paircreation1}
    \alpha_{pp} (1.022 < h\nu < 1.5\,\textrm{MeV}) = \rho [\frac{Z_{Si}^{2}}{m_{Si}} (1 - X_{IGE}) \\
    + \frac{Z_{Fe}^{2}}{m_{Fe}} X_{IGE}] (1.0063 (h\nu - 1.022\,\textrm{MeV}) \times10^{-27} cm^{2})
\end{split}
\end{equation}
\begin{equation}\label{eqn:paircreation2}
\begin{split}
\alpha_{pp} (h\nu > 1.5\,\textrm{MeV}) = 
    \rho \left[\frac{Z_{Si}^2}{m_{Si}} (1 - X_{IGE}) + \frac{Z_{Fe}^2}{m_{Fe}} X_{IGE} \right] \\
    ([0.0481 + 0.301 (h\nu - 1.5\,\textrm{MeV})] \times10^{-27} cm^{2})
\end{split}
\end{equation}
where $X_{IGE}$ is the mass fraction of the Fe group elements (IGE), and (1 - $X_{IGE}$) is the mass fraction of the intermediate mass elements (IME) and unburned carbon-oxygen.

\paragraph{Photoabsorption}

The $\gamma$-ray photons can be absorbed, and their energy is transferred to bound electrons. The effect becomes more prominent when the $\gamma$-ray energies are below 100 keV.
The photoabsorption coefficient $\alpha_{pa}$ can be calculated using equation \ref{eqn:photoabs} as given by \cite{1973AD......5...51V}, \cite{1988ApJ...325..820A} - 
\begin{equation}\label{eqn:photoabs}
\begin{split}
    \alpha_{pa}(\nu) = (1.16\times10^{-24} (\frac{h\nu}{100\,\textrm{keV}})^{-3.13} cm^{2}) \frac{\rho}{m_{Si}} (1 - X_{IGE}) \\
    + (25.7\times10^{-24} (\frac{h\nu}{100\,\textrm{keV}})^{-3} cm^{2}) \frac{\rho}{m_{Fe}} X_{IGE},
\end{split}
\end{equation}
where $\rho$ is the ejecta mass density and $X_{\rm IGE}$ is the mass fraction of Fe-group elements. 

\subsection{$\gamma$-ray packet transport}
The packet starts at its initial position with a given direction. Along the given direction, we then determine three distances - $d_{\rm interaction}$, $d_{\rm time}$, $d_{\rm boundary}$. We discuss the details of the distances in the next section (Section~\ref{event distance}). The packet is moved to the shortest distance and the specific event is handled. After the movement, the new position and direction are calculated. 
As the transport of the $\gamma$-rays is performed in the rest frame, we transform the absorption coefficients to the rest frame by using the transformation equation following Equation 2 of \cite{1972ApJ...178..779C}, and Equation 4.112 of \cite{1979rpa..book.....R}, 
\begin{equation}
    \alpha_{E} = \alpha_{E^{\prime}} (1 - \frac{\vec \mu . \vec v}{c}) 
\end{equation}
where $E$ is the packet energy in the rest frame, and $E^{\prime}$ in the co-moving frame energy. Given the position ($\vec r$), and the direction ($\vec \mu$) of the packet, we compute all the three distances and select the minimum distance to be the event reached first. The time step of the packet is increased for a packet reaching the end of the time step. For an interaction, we randomly select the type of interaction weighted by the ratio of each specific interaction absorption coefficient ($\alpha_{C}$, $\alpha_{pp}$, $\alpha_{pa}$) to the total absorption coefficient.
The packet changes its current shell if it crosses a boundary. The packet coordinates are increased by $r$ + $\vec{\mu} d$, and the time is increased by $t$ + $\frac{d}{c}$, where $d$ is the distance to the event.

\subsubsection{Distance to an event}  \label{event distance}
In the packet transport scheme we are interested in three distances - interaction distance ($d_{\rm int}$), time distance ($d_{\rm time}$), and the boundary distance ($d_{\rm boundary}$).
Knowing the total absorption and scattering coefficients ($\alpha_{total}$) due to the packets interacting with the medium, we can find the distance to an interaction ($d_{\rm int}$) event using -
\begin{equation}
    \alpha_{total}d_{int} = - ln z
\end{equation}
where $z$ is a random number between (0, 1]. The time distance ($d_{\rm time}$) is the distance the packet travels between the current and the next time step of the simulation. The distance from the inner or outer boundary ($d_{\rm boundary}$) of a shell is calculated using the current direction and location of the packet. The minimum of the two distances to a shell boundary is used to choose whether a packet crosses the inner or outer boundary. If the packet crosses the outer boundary of the last shell it is counted as an escaping packet from which the $\gamma$-ray spectrum is calculated.

\subsubsection{Interaction}
If the closest distance is $d_\textrm{interaction}$, the packet undergoes an interaction. Depending on the type of interaction the outcomes are different. 

\paragraph{Photoabsorption}

Photoabsorption is a terminal event for the high energy Monte Carlo packets as in the present version of the code we assume that the packet deposits all their energy after this interaction in-situ and instantaneously.
\paragraph{Pair production}
In our implementation within the Monte Carlo indivisible packet scheme, for a packet encountering a pair-creation event, the packet forms an electron-positron pair which deposits their kinetic energy within the ejecta and the rest of energy is carried by $\gamma$-ray packets with $\nu_{cmf}$ = 511 keV/$h$.
The $\gamma$-ray packet gets scattered in a different direction with a comoving-frame energy the same as the incident energy. In the indivisible packet scheme we do not return to the other pair \citep[see][]{2005A&A...429...19L}. 
\paragraph{Compton interaction}
A $\gamma$-ray photon with energy $E$ after scattering through an angle ($\theta$) in the plane of the scattering in the co-moving frame loses its energy by a fraction $f_{C}$, which is given by 
\begin{equation}
    f_{C} = \frac{1}{1 + x(1 - cos \theta)}
\end{equation}
and the scattering angle $\theta_{C}$ of the photon is calculated by
\begin{equation}
    \theta_{C} = cos^{-1}(1 - \frac{f_{C} - 1}{x})
\end{equation}
where $x$ = $h \nu / m_{e}c^{2}$.

A detailed discussion of our implementation of calculating the distribution of scattered photon angles can be found in Appendix~\ref{Klein-Nishina Equation}. In our simulation, knowing the Compton angle of the scattered photon, we find the new direction for the photon using the Euler-Rodrigues method. 
To rotate the co-moving direction vector $\vec{k}$ by the Compton angle $\theta$ the rotation matrix is -
\begin{align*} 
\begin{pmatrix}
a^{2} + b^{2} - c^{2} - d^{2} &  2(bc - ad) & 2(bd + ac) \\
2(bc + ad) & a^{2} + c^{2} - b^{2} - d^{2} & 2(cd - ab) \\
2(bd - ac) & 2 (cd + ab) & a^{2} + d^{2} - b^{2} - c^{2}
\end{pmatrix}
\end{align*}
where 
\begin{equation}
    a = cos (\frac{\theta}{2}); 
    b = k_{x} sin (\frac{\theta}{2});
    c = k_{y} sin(\frac{\theta}{2});
    d = k_z{2} sin(\frac{\theta}{2}); 
\end{equation}

Based on the Compton fraction ($f_{C}$) a random number is utilized to select whether the packet suffers a scattering (for $z < f_{C}$) or it loses energy to the kinetic energy of the electron. In our case, the kinetic energy of the electron is assumed to be deposited in the ejecta. In the indivisible packet scheme, the Compton scattering of the packet changes the direction and the frequency of the photons in the packet in the co-moving frame. The energy of the packet in the comoving frame is still $E_{packet, cmf}$, but the $\gamma$-ray photons now have their energy reduced by the fraction $f_{C}$. The packet properties in the rest frame are updated using the Doppler factor. 

\subsection{Spectrum and Deposition energy}
We bin the rest frame energy of all escaped packets in frequency space to form a $\gamma$-ray spectra for each individual time step. 
\begin{equation}
    L_{packet}^{\nu} = \frac{E_{\rm packet, \ rest}}{\Delta t \ \Delta \nu} 
\end{equation}

Here, $\Delta t$ is the time step width for an individual time step and $\Delta \nu$ is the frequency width of the bin. We then scale it to a flux value using -
\begin{equation}
    F_{\nu} = \frac{L_{packet}^{\nu}}{4 \pi d^{2}},
\end{equation}
where $d$ is the distance to the supernova.

The desposition energy is calculated from annihilated positrons and the $\gamma$-packets that are photoabsorbed. A schematic diagram of the code is presented in Figure~\ref{fig:figure1}.   

\section{Code comparison}  \label{code_compare}
We compare the $\gamma$-ray spectra and the energy deposited by the $\gamma$-rays and positrons in the ejecta with results from other codes. For the $\gamma$-ray spectra we used the results from \cite{2013A&A...554A..67S} and for the energy deposition we compared with results from \cite{2022A&A...668A.163B}.

\subsection{$\gamma$-ray spectra}
\tardishe takes as input the velocity, density, and mass fractions of the various elements and isotopes (we will refer to it as the ejecta model hereafter) to compute a spectrum. For most ejecta models the density and the composition profiles are defined at two specific times, which we denote as model$\_$density$\_$time, and  model$\_$isotope$\_$time.
Here, we discuss the $\gamma$-ray spectra at different epochs and compare to published spectra from the \textit{Heidelberg Supernova Model Archive} (\texttt{HESMA}) database \citep{2017MmSAI..88..312K}. For comparison, we use the model \textit{ddt N100}(\citealt{2013MNRAS.429.1156S, 2013A&A...554A..67S}), displayed in Figure~\ref{fig:figure2}. We show our spectrum calculations at around 25, 50, and 100 days for the \textit{ddt} model in Figure~\ref{fig:figure3}.
We account for the decay of the isotopes from the time the ejecta model is defined (see Figure~\ref{fig:figure2} caption) to the $t_{\rm start}$ of the simulation. In Table~\ref{tab:param_log} we list the parameters for running the models. 
The \textit{ddt N100} model by \cite{2013MNRAS.429.1156S} has a $^{56}$Ni mass of 0.6 $M_{\odot}$, but we use a subset of the isotopes and normalise. We have considered C, O, Ne (unburned elements), Mg, Si, S, Ca (IME), $^{56}$Fe, $^{56}$Co, and $^{56}$Ni (IGE). The total mass of the ejecta in this model is 1.4 $M_{\odot}$ and the total $^{56}$Ni mass present at 2 days since explosion is 0.67 $M_{\odot}$. The difference in the $^{56}$Ni mass can be seen at the late phase ($\sim$ 100 days) in the optically thin limit. The \tardishe model predicts a slightly higher flux than the comparison model by \cite{2013A&A...554A..67S}.
The spectra are calculated assuming a positronium fraction of zero (We discuss the effect of positronium in Section~\ref{positronium}). In addition to the lines due to $^{56}$Ni and $^{56}$Co the spectra has a well defined continuum at the early phases. This is mostly due to Compton scattering of the $\gamma$-ray packets. The photoelectric effect produces a low energy cut-off below 100 keV \citep{1998MNRAS.296..913G}. In the \textit{ddt} model, we notice that the $^{56}$Ni lines at 158 keV and 270 keV are present at 25 days and the line gets weaker and vanishes in the continuum at 100 days. As noted in \cite{1998MNRAS.296..913G}, \cite{2008MNRAS.385.1681S}, \cite{2013A&A...554A..67S} these two lines appear in the spectrum if there is significant $^{56}$Ni distributed to the outer layers. The cross-section of Compton scattering is higher at lower energies - so the low energy photons are down-scattered, also the higher energy photons are down-scattered to the continuum at this energy range. These lines act as a diagnostic to the explosion models (discussed further in Section~\ref{SN Ia models}). 

\begin{table}
\centering
\setlength{\tabcolsep}{3pt}
\caption{Log of the setup parameters of \texttt{TARDIS-HE}}
\label{tab:param_log}
\begin{tabular}{l c}
 \hline \hline
Parameters & Values  \\
 
\noalign{\smallskip} \hline \hline
  starting time ($t_{\rm start}$)   & 2.0 days      \\
  ending time ($t_{\rm end}$)     & 100 days        \\
  number of time steps ($t_{\rm steps}$)   & 500    \\
  time spacing ($t_{space}$)   & log                 \\
\noalign{\smallskip} \hline \hline
\end{tabular}
\end{table}
  \label{runtime_params}

\begin{figure}
\centering
\includegraphics[width=\columnwidth]{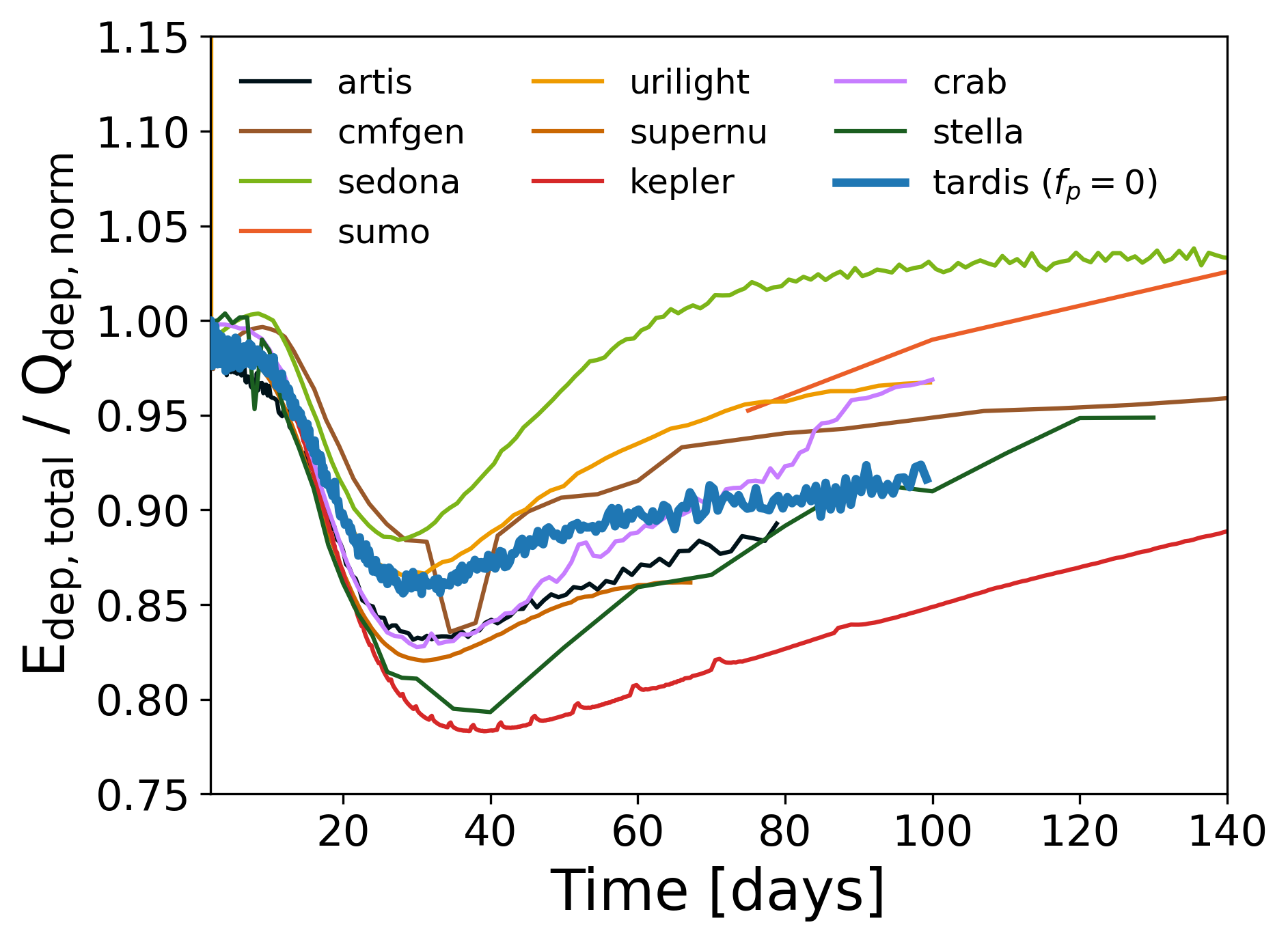}
\caption{Energy deposited by $\gamma$-rays and positrons as a function of time normalized to the analytical deposition function. We show the comparison of \tardishe with other codes. the data for the codes has been taken from \cite{2022A&A...668A.163B}.}
\label{fig:figure4}
\end{figure}

\begin{figure*}[htbp]
    \centering
    \begin{minipage}[t]{0.45\textwidth}
        \centering
        \includegraphics[width=\textwidth]{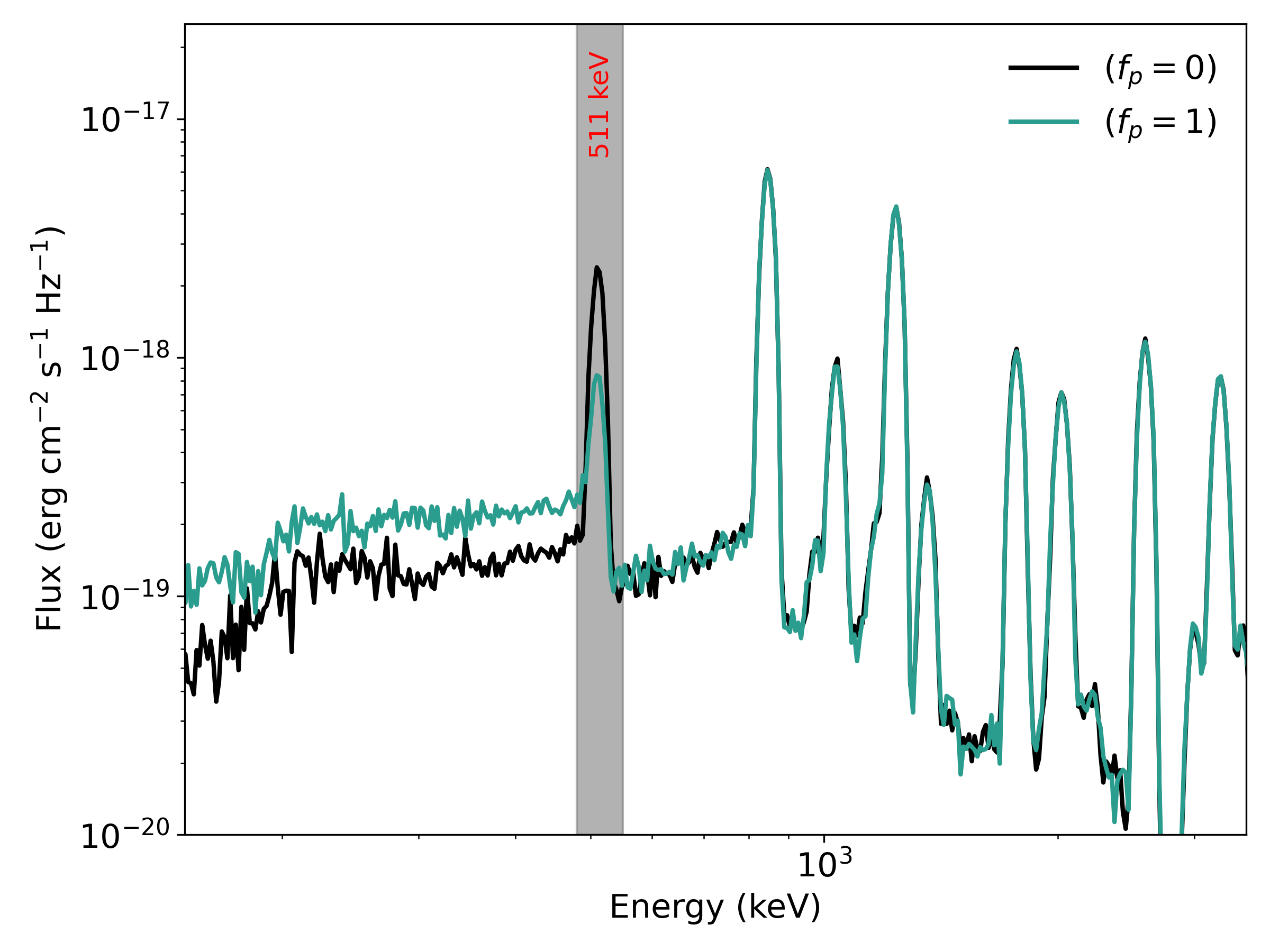}
        \caption{Comparison of the $\gamma$-ray spectra at day 100 for the \textit{ddt} model with varying positronium fraction. We use 5 $\times$ 10$^{7}$ packets in the simulation. The flux is calculated at 10 pc.}
        \label{fig:figure5}
    \end{minipage}
    \hfill 
    \begin{minipage}[t]{0.45\textwidth}
        \centering
        \includegraphics[width=\textwidth]{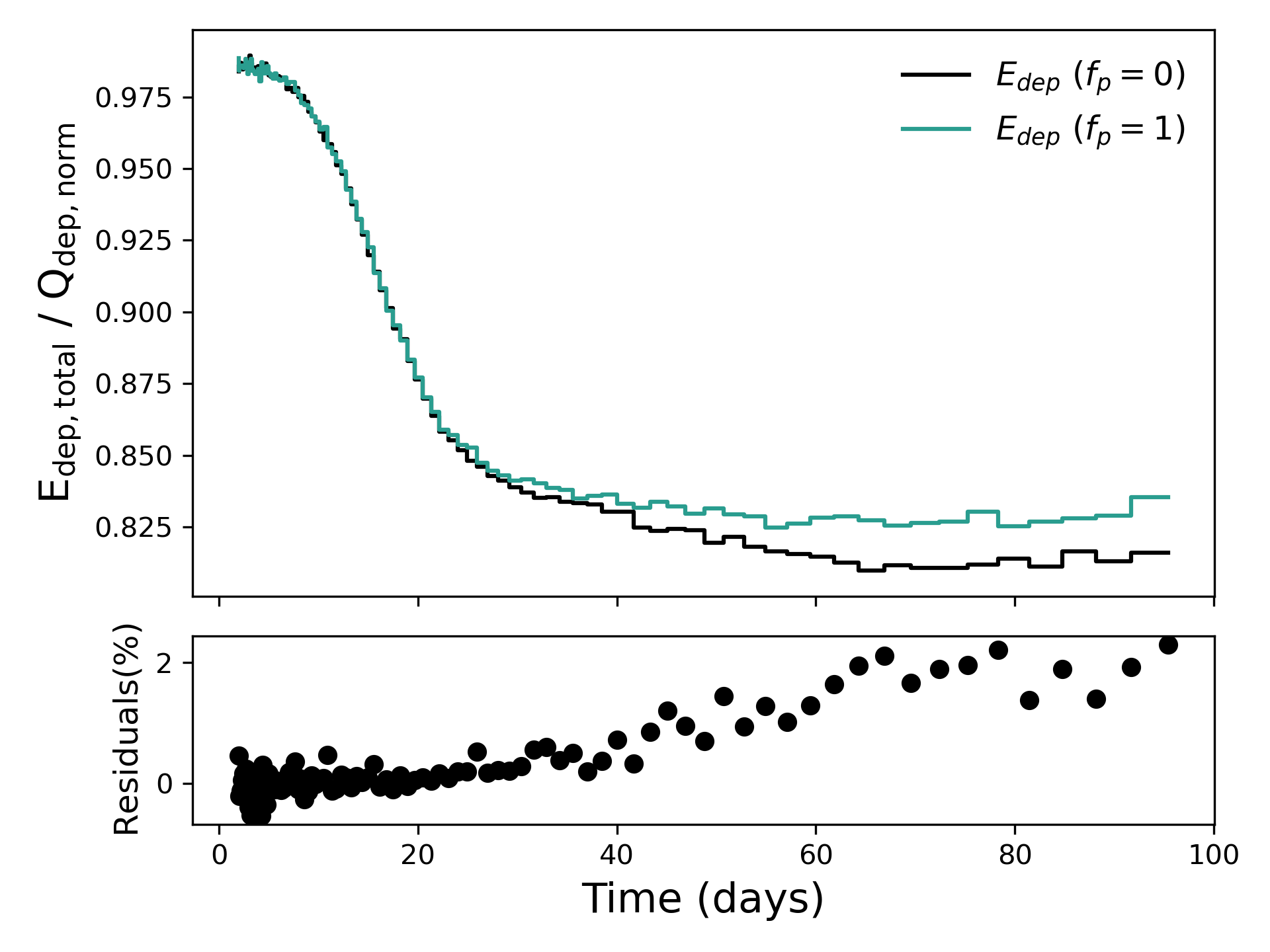}
        \caption{Comparison of the deposition function for the \textit{ddt N100} model with varying positronium fraction. We run the simulation for 500 time steps and bin the deposition energy in 100 time bins. We use 7 $\times$ 10$^{7}$ packets in the simulation.
        \label{fig:figure6}}
    \end{minipage}
\end{figure*}
\subsection{Energy deposition in the ejecta}
The $\gamma$-ray packets that are photoabsorbed deposit their energy in the ejecta, which ultimately applies to the kinetic energy of the leptons (electrons and positrons). In this work, we assume that the positrons deposit their kinetic energy in the shell where they are created. Later in the evolution of the ejecta, based on the amount of material and the ejecta composition, positron transport becomes important \citep{1999ApJS..124..503M}. We restrict our energy deposition calculations up to 100 days in this study. To validate the deposition by the gamma rays and positrons as a function of time, we compare \tardishe with other codes in the literature \citep[see][for an overview of other codes]{2022A&A...668A.163B}. We use the \textit{toy06} model for normal Ia from \cite{2022A&A...668A.163B} for the comparison. To highlight the differences, we normalize with respect to an analytical deposition function given by
\begin{equation}
\begin{split}
    Q_{dep, norm} = \frac{M_{Ni}}{M_{\odot}}(0.97[1 - e^{-(40d/t)^{2}}] + 0.03) \\
    \times (6.45 e^{-t/8.8d} + 1.45 e^{-t/111.3d}) \times 10^{43} erg~s^{-1},
\end{split}
\end{equation}
where $M_{\rm Ni}$ = 0.6 $M_{\odot}$ is the $^{56}$Ni mass in the \textit{toy06} model at 2.0 days. In the first term, a fraction 0.03 of the energy is in the kinetic energy of the positrons, while the rest is in $\gamma$-rays with a purely absorptive optical depth given by $\tau_{\gamma} = \frac{t_{0}^{2}}{t^{2}}$ with $t_{0}$ = 40~d being the $\gamma$-ray escape time. The second term is similar to Equation~19 of \cite{1994ApJS...92..527N}.
In the calculation of our \textit{deposition curve} we used 5 $\times$ 10$^{7}$ packets with 500 time steps between 2 and 100 days for a positronium fraction of zero. We use \textit{logarithmic} time steps to capture the evolution of the deposition energy. See also Appendix~\ref{time convergence} for convergence of the energy deposition function with time-steps in our implementation.

We compare the deposition energy obtained with \tardishe with that from other Monte Carlo codes like \texttt{ARTIS} \citep{2007MNRAS.375..154S}, \texttt{SEDONA} \citep{2006ApJ...651..366K}, \texttt{SUMO} (\citealt{2011A&A...530A..45J}, \citealt{2012A&A...546A..28J}), \texttt{URILIGHT} (\citealt{2019MNRAS.484.3951W, 2022ascl.soft09012E}), the Monte Carlo transport of gamma-ray photons in \texttt{CMFGEN} \citep{1998ApJ...496..407H, 2012MNRAS.424..252H}. We find that even though the codes predict similar deposition energy in the initial phases there are differences in the later times as the SN evolves. These differences are due to (but not limited to) the nuclear decay data, the opacity approximations used, and the time sampling of the Monte Carlo packets. Some of the codes use a constant grey opacity for $\gamma$-rays. 
We also show other radiative transfer codes that use non Monte Carlo transport methods like, \texttt{CRAB} \citep{2004AstL...30..293U}, \texttt{KEPLER} \citep{1978ApJ...225.1021W}, \texttt{SUPERNU} (\citealt{2013ApJS..209...36W, 2014ApJS..214...28W}), and \texttt{STELLA} \citep{1993A&A...273..106B}.

\section{Results} \label{results}   

We discuss the effect of positronium on the $\gamma$-ray spectra and energy deposition function and employ \tardishe to SNe Ia models to distinguish between them using $\gamma$-ray features. We also study the effect of positronium contribution on the line ratio for the SNe Ia models.

\subsection{Positronium contribution} \label{positronium}
The decay of $^{56}$Co is a source of positrons (19 $\%$ of $^{56}$Co decay produces positrons). For the case of a positronium fraction of 1, the 3$\gamma$ decay contributes to the continuum below 511 keV (see Equation~\ref{Ore and Powell}). We consider the \textit{ddt N100} model and measured the flux of the 511 keV line after subtracting the continuum. 
We measure line and continuum fluxes using the \texttt{specutils} \citep{nicholas_earl_2024_14042033} package in python. The line flux is measured by integrating over a one-dimensional Gaussian function fit to the line. 
There is a decrease in flux in the 511 keV line by $\sim$ 70 $\%$ at around 95 days for a positronium fraction of 1. The flux in the continuum region below the 511 keV line increases for the same case. Figure~\ref{fig:figure5} shows the spectra for the \textit{ddt} model with varying positronium fraction. 

Next, we show the effect of changing the positronium fraction parameter on the energy deposition curve for the \textit{ddt} model in Figure~\ref{fig:figure6}. For a positronium fraction of 1, there is up to 2 $\%$ increase in the energy deposition in the later phases. The photoabsorption opacity is mainly important for the low energy continuum, while the Compton opacity is dominant within the energy range 100 -- 1000 keV \citep[See Figure~1 of][]{1995ApJ...446..766S}. 
The fact that more photons in the continuum are close to the photoabsorption limit and hence have a higher probability of contributing to the opacity affects the energy deposition by the $\gamma$-rays. Also, the 511 keV $\gamma$-ray line is produced by the positron annihilation of $^{56}$Co. In the early phase ($\sim$ 25 days) the strength of the 511 keV line is weaker due to the fact that $^{56}$Co has a half-life of around 77 days and is still decaying in the ejecta, and due to higher density the continuum opacity is higher. The 511 keV line get stronger as the ejecta expands and the continnum opacity decreases, so the effect of positronium ($f_{p}$=1) can be observed. The detection of the ortho-Ps continuum in $\gamma$-ray observations is a definite proof of positronium formation.

\begin{figure}
\centering
\includegraphics[width=\columnwidth]{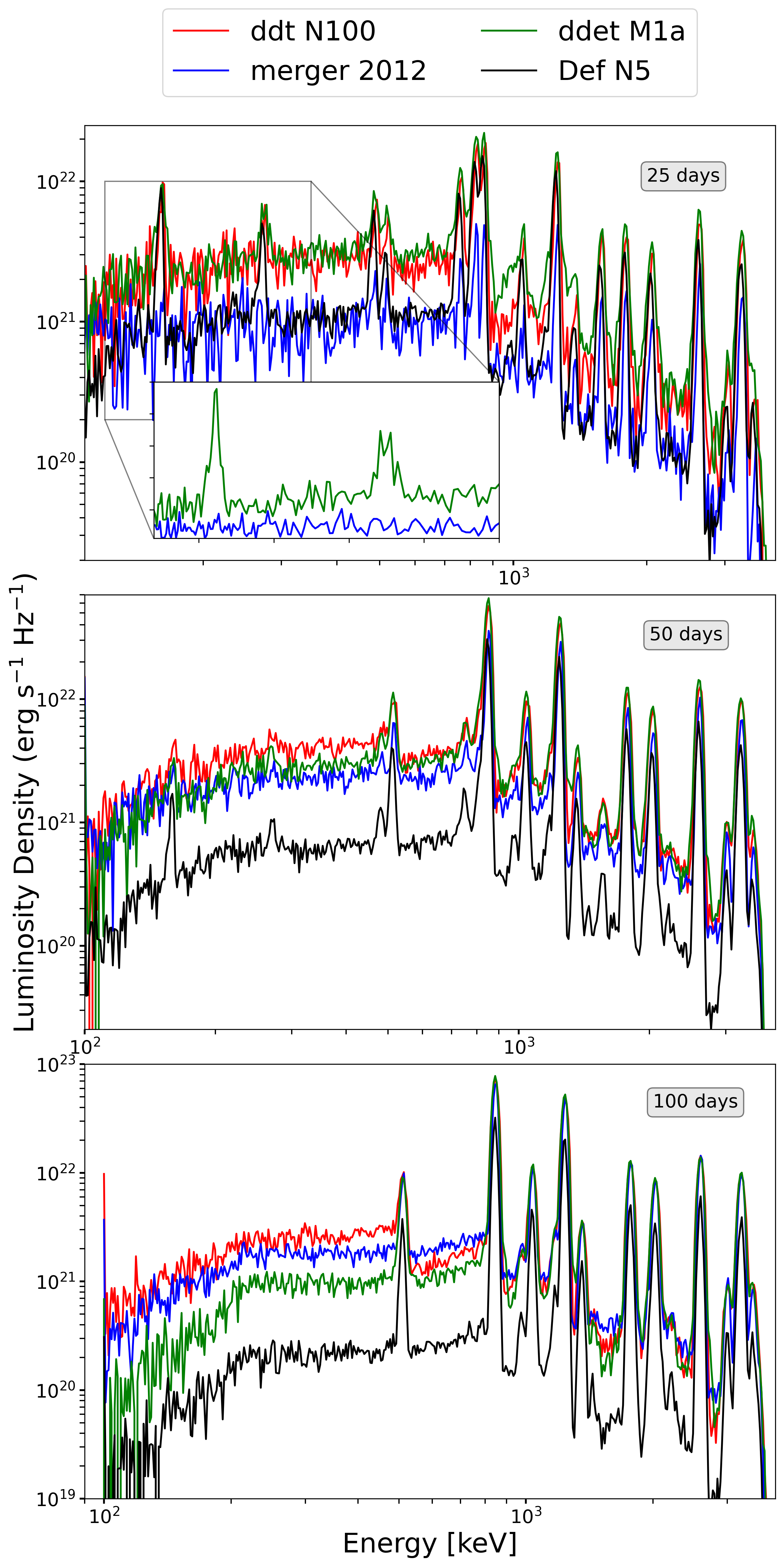}
\caption{$\gamma$-ray spectra generated with \tardishe at 25, 50, and 100 days since explosion for four different SNe Ia models. All the models are run with 7 $\times$ 10$^{7}$ packets with an $f_{p}$=1. The zoomed inset shows the region between 120 and 350 keV including the 158 and 270 keV lines of $^{56}$Ni for the \textit{ddet M1a} and the \textit{merger} models.}
\label{fig:figure7}
\end{figure}

\subsection{SN Ia models}  \label{SN Ia models}
 We employ \tardishe to four different Type Ia models. We consider the evolution of the 511 keV annihilation line and compare it with the 1238 keV line of $^{56}$Co to find the effect of positronium formation. We consider the angle averaged density profiles and compositions from the \texttt{HESMA} database. For the simulations $^{56}$Ni, $^{56}$Co, C, O, Ne, Mg, Si, S, Ca and Fe are considered if not otherwise mentioned. Since we are taking a subset of elements from the models wherever applicable the mass fractions are normalized to 1. 

\begin{figure*}
\centering
\includegraphics[width=0.8\linewidth]{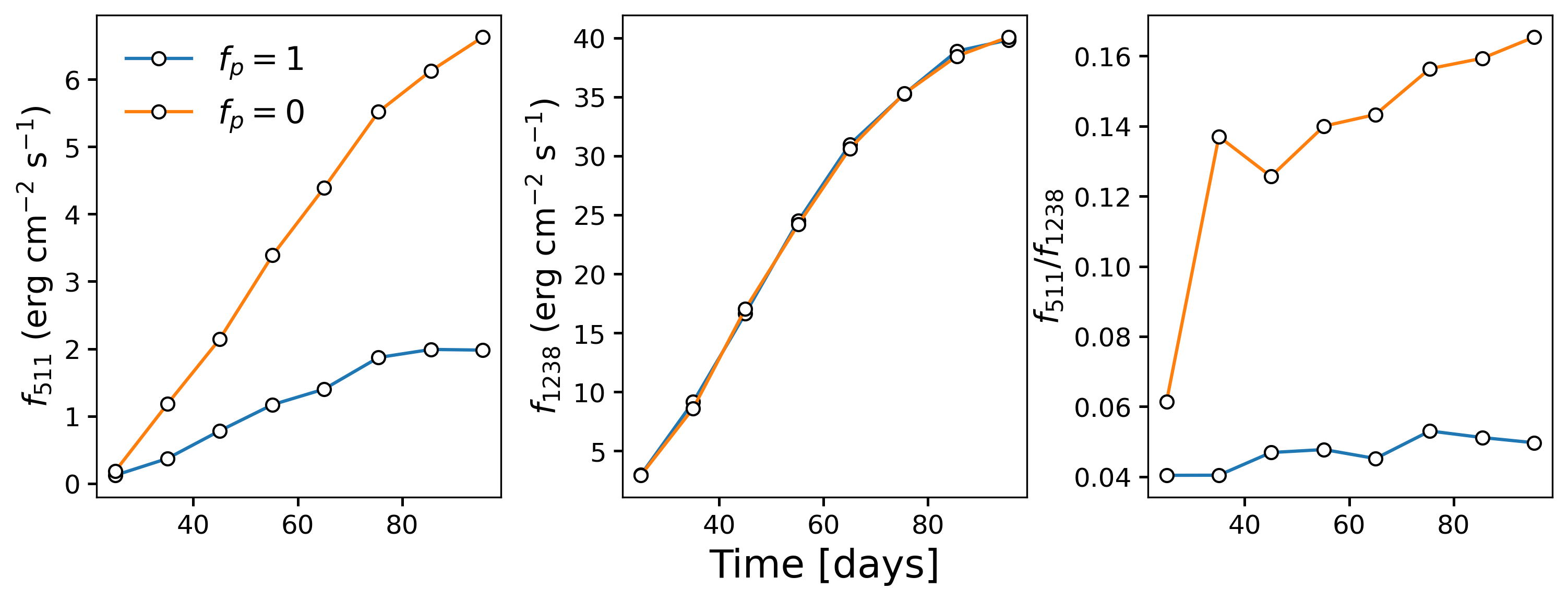}
\caption{Evolution of the (a) 511 keV line (b) 1238 keV line flux of $^{56}$Co and the ratio of these two line fluxes for different positronium fractions for the violent merger model. The flux is calculated at 10 pc.}
\label{fig:figure8}
\end{figure*}

\textit{WD merger --} This model considers the \textit{violent merger} of two white dwarfs of masses 0.9 and 1.1 $M_{\odot}$ \citep{2012ApJ...747L..10P}. The material from the secondary is accreted onto the primary. The accreted materials are compressed on the surface of the primary and heated up. As a result, hotspots are formed where carbon is ignited. The conditions for prompt detonation are reached and the system explodes. The $^{56}$Ni mass produced is 0.67 $M_{\odot}$, and the ejected mass is 1.9 $M_{\odot}$. In the three-dimensional simulation by \cite{2012ApJ...747L..10P} the innermost part of the ejecta is devoid of Fe-group elements and $^{56}$Ni and consist mainly of unburned and intermediate elements. We call \textit{merger 2012} hereafter.

\textit{Delayed detonation --} The \textit{ddt N100} model is based on non-rotating isothermal WD in hydrostatic equilibrium composed of C, O, and Ne. In \cite{2013MNRAS.429.1156S} different explosion setups are explored based on multi-spot ignition. This particular model has 100 ignition spots around the center of the WD. An initial subsonic deflagration turns to a supersonic detonation and hence unbinds the WD. The $^{56}$Ni produced is 0.67 $M_{\odot}$ and the ejected mass is 1.4 $M_{\odot}$. In this model, the central part of the ejecta is dominated by $^{56}$Ni and Fe group elements.

\textit{Deflagration -- } We consider the \textit{def N5} model which is based on the WD configuration from \cite{2013MNRAS.429.1156S}, but in this model it is assumed that no delayed detonation occurs. Hence, these models depict scenarios where only pure deflagration works and some portion of the WD is still bound. In the specific model \textit{N5} from \cite{2014MNRAS.438.1762F} the deflagration is set with five ignition points. The $^{56}$Ni mass produced in this model is 0.17 $M_{\odot}$. The composition is mixed with $^{56}$Ni present throughout the ejecta.

\textit{Double detonation--} This model considers a WD with a carbon-oxygen core and a He shell. In the model \textit{ddet M1a} from \cite{2020A&A...635A.169G} the total mass of the WD is 1.05 $M_{\odot}$ and an initial ignition in the He shell leads to a second detonation in the CO core. As a result of He shell burning there is $^{56}$Ni in the outer layers. The CO burning produces $^{56}$Ni mostly in the inner layers.  In this model we also consider $^{44}$Ti, $^{48}$Cr, and $^{52}$Fe in the composition.

In Figure~\ref{fig:figure7}, we show the spectral evolution for the above models to find distinguishing signatures that can reveal the explosion model. In the early phase, the spectrum has a continuum formed by the Compton scattering of photons along with line emission from $^{56}$Ni and $^{56}$Co. Since the compton opacity is sensitive to the electron number density ($n_{e}$), the continuum decreases as the SN evolves. This lead to increasing strength of the lines with respect to the continuum at later epochs. The compton opacity decreases with energy and hence the low energy photons are scattered more as compared to the high energy photons. In the optically thin limit, the line flux gives a direct estimation of the mass of radiaoctive isotopes.

The \textit{merger 2012} has more material (IME-rich) above the $^{56}$Ni-rich region than other models. The lines 158 keV and 270 keV of $^{56}$Ni which are present in \textit{ddt N100}, \textit{ddet M1a}, and \textit{def N5} are lost in the continuum in \textit{merger 2012} as the optical depth to compton scattering is higher in the later (see inset of the top panel of Figure~\ref{fig:figure7}). In the case of \textit{ddet M1a} even if it has lower $^{56}$Ni (0.61 $M_{\odot}$) mass than \textit{merger 2012} (0.67 $M_{\odot}$) and \textit{ddt N100} (0.67 $M_{\odot}$) the outer layers of the model have significant $^{56}$Ni at lower optical depths. Hence, more $\gamma$-rays can escape the outermost layer of the ejecta. As a result the flux of the lines with respect to the continuum are higher in this case. The \textit{def N5} model has low $^{56}$Ni mass (0.17 $M_{\odot}$) and a lower flux than the model \textit{ ddet M1a} and the \textit{ddt N100}. But its early phase continuum flux is similar to \textit{merger 2012} which indicates that even if \textit{merger 2012}  has more $^{56}$Ni mass, the column density of electrons is larger leading to a similar continuum compared to the \textit{def N5} model. However, the flux decreases in the optically thin limit owing to less $^{56}$Ni mass in \textit{def N5}.

The line ratio of two similar isotopes at different energies is dependent on the opacity at those energies at early times and become similar in the optically thin limit (mostly independent of the model composition). We choose the 511 keV line and compare it with a relatively strong and unblended line of $^{56}$Co at a different energy. The line flux for the 511 keV line is measured by subtracting the continuum, and for the 1238 keV line we do not perform continuum subtraction. 

\cite{2008MNRAS.385.1681S} studied line ratios of similar and different isotopes at different energy ranges. They also studied the hardness ratios at different energies for different toy supernova models to identify distinguishing features in the $\gamma$-ray spectra. In Figure~\ref{fig:figure8}, we show the fluxes for the 511 keV and 1238 keV lines and their ratio.
The line flux of the 511 keV and 1238 keV lines increases as the opacity decreases (see Figure~\ref{fig:figure8}a, and \ref{fig:figure8}b). With a positronium fraction of 1, the flux of the 511 keV line decreases but that of 1238 keV line remains the same. The formation of positronium reduces the flux of the 511 keV line as it is distributed to the continuum with a probability of 75 $\%$ in 3$\gamma$-decay. For all the models considered in the work, the line ratio of 511 keV to 1238 keV decreases compared to the case when no positronium is formed. The ortho-Ps continuum is independent of the composition of the ejecta unlike the Compton scattered continuum.

\section{Conclusion} \label{conclusion}

We study the effect of positronium formation on the $\gamma$-ray spectra and the energy deposition function of SNe Ia. For this purpose, we developed a new module named \tardishe for time-dependent three-dimensional $\gamma$-ray transport for the radiative transfer code \textsc{tardis}. \tardishe can handle any decay chain without metastable states. The code takes a one-dimensional ejecta model for a SN and provides $\gamma$-ray spectra and light curves. Approximations to opacities and new microphysics can be added to explain observables. The code provides time-dependent $\gamma$-ray opacities and energy deposition which can be further used to calculate UV-Optical-IR emissions in radioactively powered transients.  

For numerical implementation in the code, we quantify the fraction of positrons forming positronium before annihilating by the positronium fraction. We varied the fraction between the limits of positrons forming no positronium and forming positronium with a probability of 100$\%$. The formation and decay of positronium depend on the positron motion in the ejecta and electron number density. The positrons travel a longer distance in the ejecta before annihilating at later times in the supernova evolution. The temperature and the ionization state of the medium play an important role in whether a positronium will decay by producing 2$\gamma$ or 3$\gamma$ \citep{1976ApJ...210..582C, 1987ApJ...323..159L}. The collisions of positronium with external electrons and positrons can lead to their dissociation by the pick-up annihilation process.
More collisions within the ejecta will lead to suppression of ortho-Ps as the timescale for a positronium to decay will be comparable to or lower than ortho-Ps decay time. There is also flipping of the triplet state to the singlet state by collisions with electrons, so the flux contributed to the continuum by the 3$\gamma$ decay changes. A high positronium fraction will mean that the positrons tend to be produced in regions of low ionization fraction \citep[see Figure 26 of][]{2011RvMP...83.1001P}.

In our simplified parameterized approach we kept the positronium fraction fixed, but it changes over time as the ejecta evolves. We find that for positronium forming with 100 $\%$ probability and 75 $\%$ of the positronium decaying through 3$\gamma$ the flux in the 511 keV line decreases $\sim$ 70 $\%$ at $\sim$ 100 days, and the energy deposition increases by up to 2 $\%$ around 100 days than when no positronium is formed. The line flux ratio of 511 keV to 1238 keV decreases for all the models as a result of this. Observations of late phase $\gamma$-ray spectra of nearby SNe Ia may constrain positronium formation. In the late phase of the supernova evolution, the non-thermal electron energy from the $\gamma$-rays and positrons goes into Coulomb heating, ionization, excitation and bremsstrahlung (via the Spencer-Fano formalism). This impacts the UV, optical and infrared spectra but has negligible impact on the role of positronium in 511 keV emission as these secondary processes produce only lower energy photons.

We consider the time evolution of the $\gamma$-ray spectra for different SNe Ia models. The low-energy $^{56}$Ni lines and the evolution of the continuum can be used to distinguish between the explosion models. Although SNe Ia are mainly powered by the radiaoctive decay of $^{56}$Ni, at later times other decay chains become important, e.g. $^{55}$Co $\xrightarrow{\text{0.73~d}}$ $^{55}$Fe$\xrightarrow{\text{1000.2~d}}$ $^{55}$Mn, $^{57}$Co $\xrightarrow{\text{272~d}}$ $^{57}$Fe \citep[see][]{2014MNRAS.441.3249D, 2016ApJ...819...31G}. Many SNe Ia models exist, which have shown that apart from $^{56}$Ni there are other isotopes like $^{52}$Fe or $^{48}$Cr which are formed in the outer layers \citep{2017MNRAS.472.2787N, 2020A&A...642A.189M, 2021MNRAS.502.3533M}. The resulting effect is seen in the early UV-optical light curves of SNe Ia. Our results show that while the impact of positronium formation on $\gamma$-ray spectra and energy deposition is modest, it introduces measurable differences, highlighting the need to account for it in precise modeling of SNe Ia.

\section{Acknowledgments}
We thank the referee for providing constructive comments. We acknowledge the \texttt{HESMA} database \citep{2017MmSAI..88..312K}. AD acknowledges Jing Lu and Joshua V. Shields for useful discussions. AF acknowledges support by the European Research Council (ERC) under the European Union's Horizon 2020 research and innovation programme (ERC Advanced Grant KILONOVA No.~885281), the Deutsche Forschungsgemeinschaft (DFG, German Research Foundation) - Project-ID 279384907 - SFB 1245, and MA 4248/3-1, and the State of Hesse within the Cluster Project ELEMENTS. 
AGF acknowledges support by the National Science Foundation (NSF), grant number 2311323, and NASA HST-AR-16613.002-A. WEK acknowledges funding support by NSF grant number OAC-2311323, AAG-2206523, and NASA grants HST-AR-16613.002-A, HST-GO-16885.011-A. AD acknowledges funding support from HST-AR-16613.002-A and HST-GO-16885.011-A.
SAS acknowledges funding from UKRI STFC grant
ST/X00094X/1.

Contributions: \textit{Conceptualization} - Stuart A Sim, Andrew Fullard, Wolfgang E Kerzendorf, and Anirban Dutta; \textit{Data Curation} - Andrew Fullard; \textit{Formal Analysis} - Anirban Dutta and Andrew Fullard; \textit{Funding Acquisition} - Andrew Fullard, Wolfgang E Kerzendorf, Or Graur, and Saurabh Jha; \textit{Investigation} - Anirban Dutta, Andrew Fullard, and Wolfgang E Kerzendorf; \textit{Methodology} - Anirban Dutta, Andrew Fullard, and Wolfgang E Kerzendorf, J T O'Brien; \textit{Project Administration} -  Wolfgang E Kerzendorf; \textit{Software} - Anirban Dutta, Andrew Fullard, Andreas Flörs, and Cecelia Powers; \textit{Resources} - Wolfgang E Kerzendorf; \textit{Supervision} - Wolfgang E Kerzendorf; \textit{Validation} - Anirban Dutta; \textit{Visualization} - Anirban Dutta; \textit{Writing-original draft} - Anirban Dutta, Wolfgang E Kerzendorf; \textit{Writing-review and editing} - Anirban Dutta, Wolfgang E Kerzendorf, Andrew Fullard, Or Graur, J T O'Brien, Stuart A Sim, Andreas Flörs.

Software: \textsc{radioactivedecay} \citep{Malins2022}, \textsc{Astropy} \citep{2013A&A...558A..33A, 2018AJ....156..123A, 2022ApJ...935..167A} \textsc{numpy} \citep{harris2020array}, \textsc{pandas} \citep{mckinney-proc-scipy-2010, reback2020pandas}, \textsc{matplotlib} \citep{Hunter:2007}, \textsc{tardis} \citep{2014MNRAS.440..387K}, \textsc{specutils} \citep{nicholas_earl_2024_14042033}.

\appendix

\section{Sampling the three photon decay of positronium} \label{three photon decay}

\begin{figure}
\centering
\includegraphics[width=0.5\columnwidth]{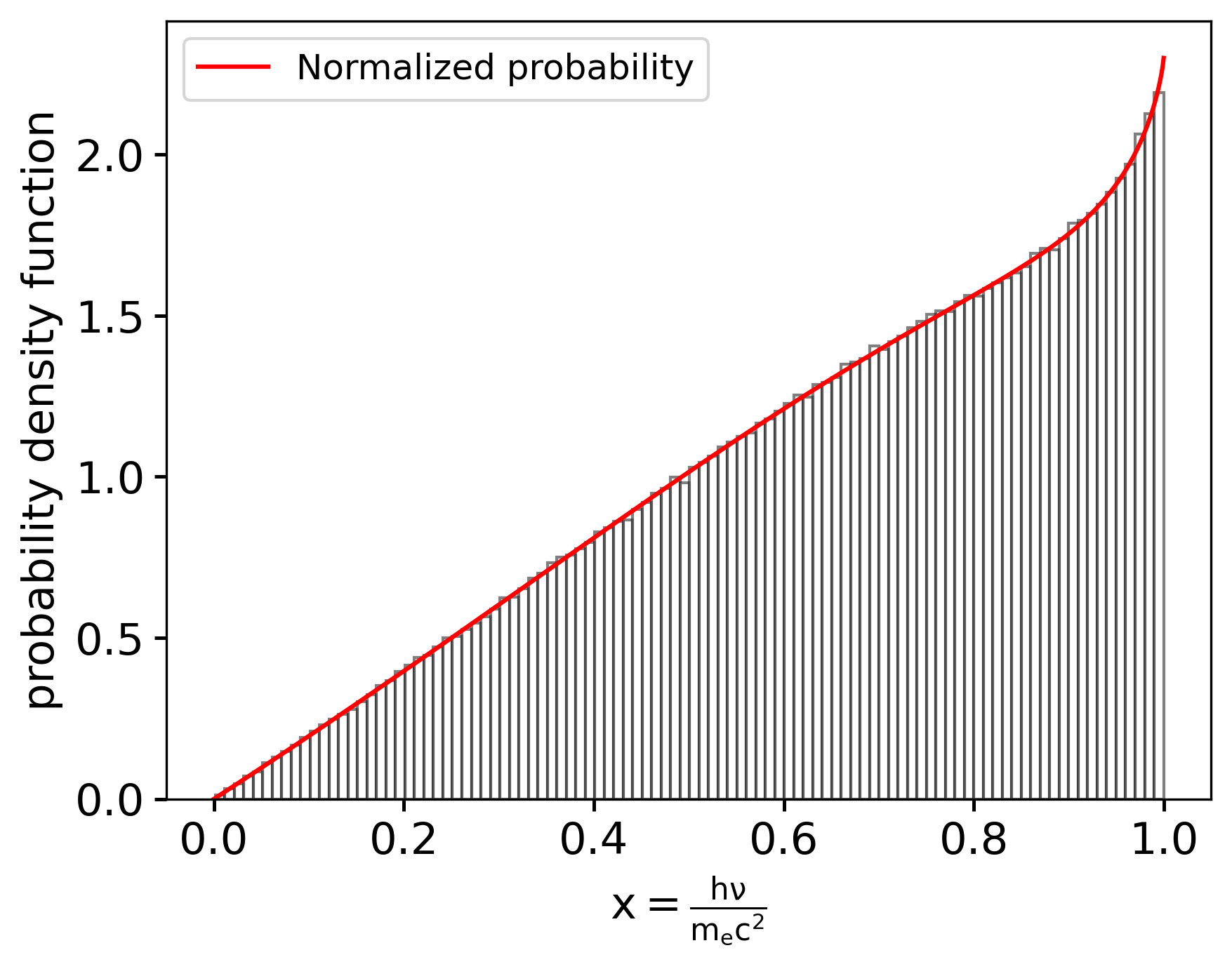}
\caption{The sampling of the photon energies for the case of three photon decay as given by equation \ref{Ore and Powell}. In this example we draw 1 $\times$ 10$^{6}$ photon energies and uniformly binned in 100 bins. The \textit{abscissa} is the photon energy in terms of electron rest mass energy. }
\label{fig:figure9}
\end{figure}

The energy of the photons in the three-photon decay is sampled using the probability distribution function (PDF) from Ore and Powell (1949) shown in Figure~\ref{fig:figure9}
\begin{equation} \label{Ore and Powell}
\begin{split}
F(x) = 2 [\frac{x(1 - x)}{(2 - x)^{2}} - \frac{2(1 - x)^{2}}{(2 - x)^{3}}log(1 - x) \\
+ \frac{2 - x}{x} + \frac{2(1 - x)}{x^{2}}log(1 - x)],
\end{split}
\end{equation}
where $x$ is defined as $\frac{h\nu}{m_{e}c^{2}}$. 
The cumulative distribution function (CDF) is found from the normalized PDF using 
\begin{equation}
    F_{CDF}(u) = \int_{0}^{u} F(u^{\prime}) du^{\prime}
\end{equation}
Then the inverted CDF $F_{CDF}^{-1}(z)$ for random values $z$ within (0, 1)  gives the distribution of photon energies which satisfies the given $F(x)$.

\begin{figure}
\centering
\includegraphics[width=0.8\columnwidth]{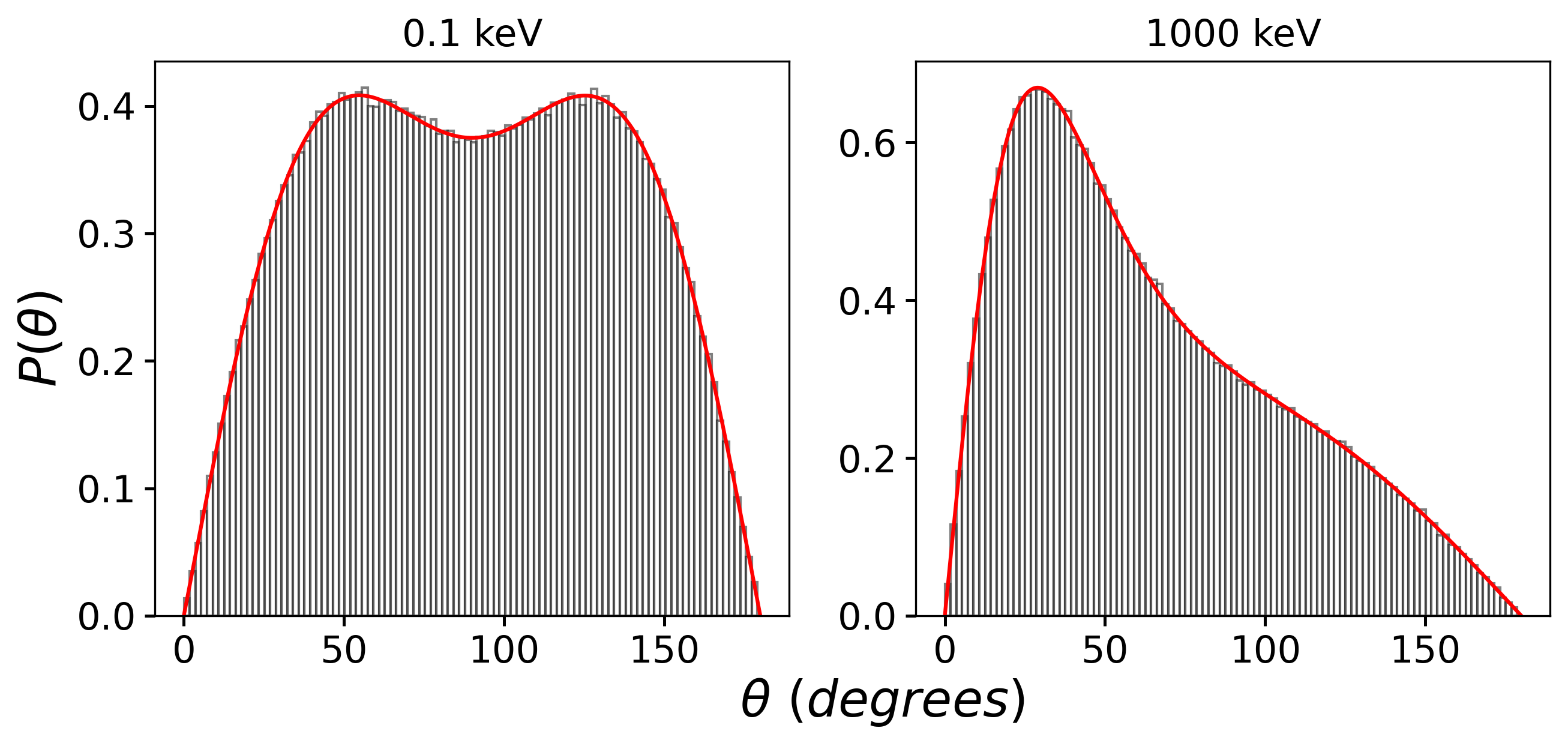}
\caption{Distribution of scattered photon angles. The red curve shows the scattering angle distribution as predicted by the Klein-Nishina equation.}
\label{fig:figure10}
\end{figure}

\section{Klein-Nishina equation} \label{Klein-Nishina Equation}

The differential cross-section of scattering of a photon with a free electron assuming isotropy in the azimuthal direction is given by the Klein-Nishina formula \citep[see Equation 8.7.41 of][]{1995qtf..book.....W}-
\begin{equation} \label{KN equation}
    \frac{d\sigma}{d\Omega} = \frac{1}{2}r_{e}^{2}f_{C}^{2} [f_{C} + f_{C}^{-1} - sin^{2}(\theta)]
\end{equation}
where $r_{e}$ is the classical electron radius, and $\theta$ is the angle between the incident and scattered photon direction.
Integrating the equation from 0 to $\theta$, we get the partial cross-section up to an angle $\theta$ given by -
\begin{equation}    
\begin{split}
    \sigma(f_{C})_{partial} = \frac{3\sigma_{T}}{8x}(\frac{(x^{2} - 2x - 2)ln(f_{C})}{x^{2}} \\
    + \frac{f_{C}^{2} - 1}{2f_{C}^{2}} + \frac{f_{C} - 1}{x} [\frac{1}{x} + \frac{2}{f_{C}} \\
    + \frac{1}{xf_{C}}])
\end{split}
\end{equation}
$\sigma_T$ is the Thomson scattering cross-section, and $x = $ $h\nu/m_e c^2$ where $\nu$ is the packet frequency.
We use the partial cross-section as given in the \texttt{ARTIS} \footnote{\url{https://github.com/artis-mcrt/artis}} code \citep{2007MNRAS.375..154S, 2009MNRAS.398.1809K}.

In \tardishe, we find the scattering angle ($\theta_{C}$), and the fraction of energy lost ($f_{C}$), by solving the equation 
\begin{equation}
    \sigma_{partial} / \sigma_{total} = z,
\end{equation}
by a bisection method. Here $z$ is a random number between [0, 1). 
The scattering angle of the packet is found from -
\begin{equation}
    \theta_{C} = cos^{-1}(1 - \frac{f_{C} - 1}{x})
\end{equation}
Our implementation of calculating the Compton scattering can be tested against the theoretical calculation. 
The probability distribution of the scattered photon between angle $\theta$ and $\theta + d \theta$ can be obtained from Equation~\ref{KN equation} as -
\begin{equation}
    P(\theta) = \frac{1}{\sigma} \frac{d\sigma}{d\Omega} 2 \pi sin (\theta)
\end{equation}
where the total scattering cross-section can be found by integrating the distribution over $\phi$ (0 and 2$\pi$) and $\theta$ (0 and $\pi$). In Figure~\ref{fig:figure10}, we show the scattered angle distribution of our implementation with the predicted pdf. The distribution is shown for 1 $\times$ 10$^{6}$ photons for incident energies 0.1 keV and 1000 keV. In the lower energy limit ($E << m_{e}c^{2}$), the distribution tends to the Thomson cross-section as the photons scatter forward and backward. But in the higher energy range ($E >> m_{e}c^{2}$), the photons are scattered at small angles as the scattering cross-section decreases with energy.

\section{Convergence of the energy deposition} \label{time convergence}

\begin{figure}
\centering
\includegraphics[width=0.5\columnwidth]{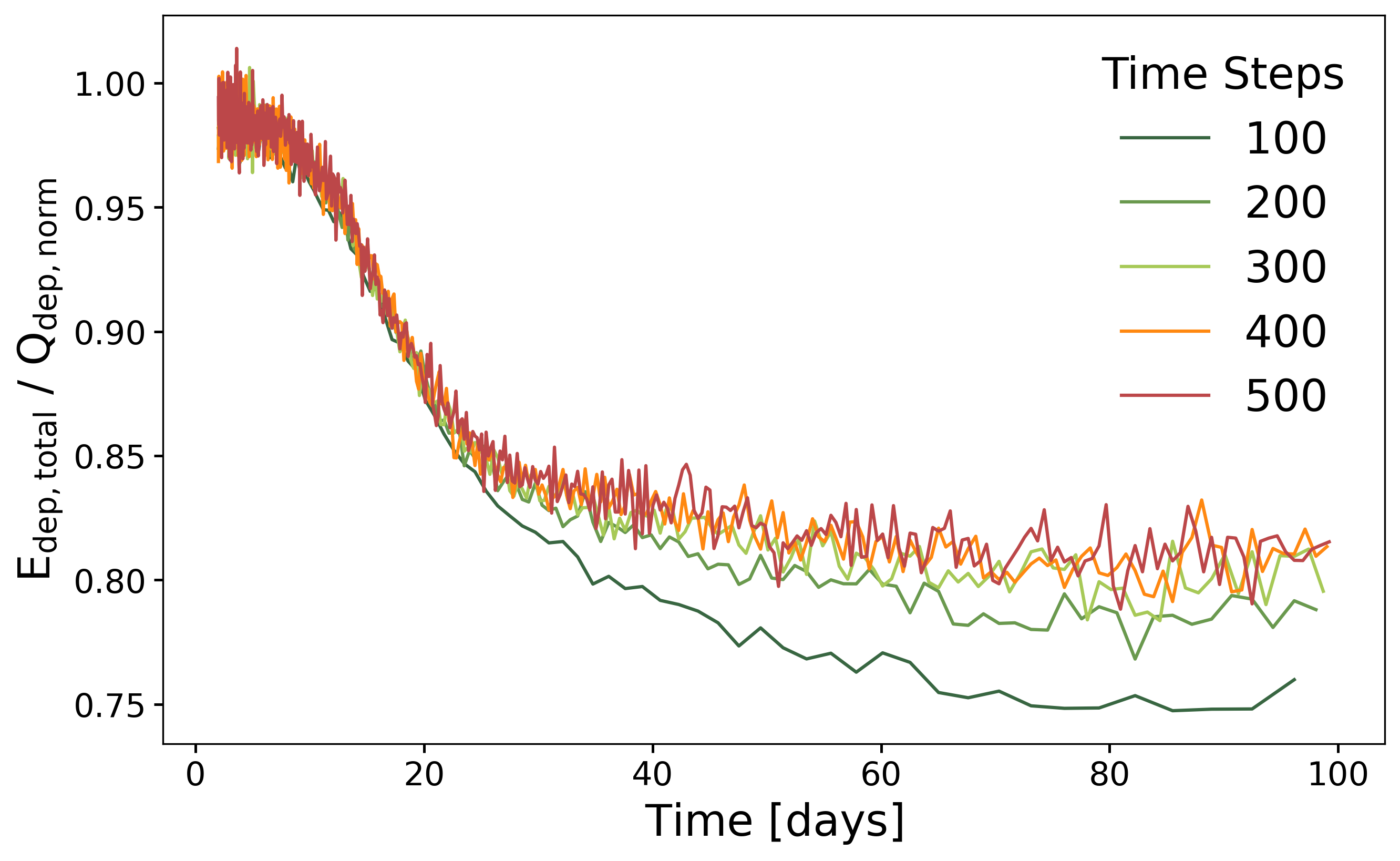}
\caption{The energy deposited by the $\gamma$-rays and positrons normalized by the analytic deposition function for various time steps.}
\label{fig:figure11}
\end{figure}

We varied the time steps in our simulations from 100 to 500 and find that the change in deposition energy with time reaches a convergence for over 400 time steps (see Figure~\ref{fig:figure11}). To capture the decay of $^{56}$Ni and $^{56}$Co which have half-lives of $\sim$6 and $\sim$77 days respectively our time resolution is less than 0.1\ day for less than 10 days and increases to 0.8\ day for around 100 days. 

\bibliography{tardis_gamma_ray}{}
\bibliographystyle{aasjournal}

\end{document}